\documentclass[aps,preprint,groupedaddress,showpacs]{revtex4}
\usepackage{amsmath}
\usepackage{graphicx}
\begin{document}

\title{The Higgs boson: from the lattice to LHC}

\author{P. Cea$^{1,2}$}
\email[]{Paolo.Cea@ba.infn.it}
\author{L. Cosmai$^{1}$}
\email[]{Leonardo.Cosmai@ba.infn.it}
\affiliation{$^1$INFN - Sezione di Bari, I-70126 Bari, Italy\\
$^2$Physics Department, Univ. of Bari, I-70126 Bari, Italy }

\begin{abstract}
We discuss the triviality and spontaneous symmetry breaking scenario where the Higgs boson without self-interaction coexists with spontaneous symmetry breaking. We argue that non perturbative lattice investigations support this scenario. Moreover, from lattice simulations we predict that the Higgs boson is rather heavy. We estimate  the Higgs boson mass  $m_H = 754 \,  \pm \, 20 \, {\text{(stat)}}  \, \pm \, 20 \, {\text{(syst)}} \,  {\text{GeV}}$ and the Higgs total width 
$\Gamma( H)  \, \simeq \, 340 \, {\text{GeV}}$.
\end{abstract}


\maketitle

\section{Introduction}
\label{Introduction}
A cornerstone of the Standard Model is the mechanism of spontaneous symmetry breaking that, as is well known, is mediated by the Higgs boson. Then, the discovery of the Higgs boson is the highest priority of the Large Hadron Collider (LHC)~\cite{Higgs_book,Djouadi:2005gi}. \\
\indent
Usually the spontaneous symmetry breaking in the Standard Model is implemented within the perturbation theory which leads  to predict that the Higgs boson mass squared, $m^2_H$, is proportional to $\lambda_R \;  v^2_R$, where $v_R$ is the known weak scale (246~GeV) and $\lambda_R$ is the renormalized scalar self-coupling.  However, it has been  conjectured since long time ~\cite{Wilson:1973jj} that self-interacting four dimensional scalar field theories are trivial, namely $\lambda_R   \rightarrow 0$ when $\Lambda \rightarrow \infty$  ($\Lambda$ ultraviolet cutoff). Even though no rigorous proof of triviality exists, there exist several results which leave little doubt on the triviality conjecture~\cite{Luscher:1987ay,Luscher:1988ek,Sokal_book,Lang:1993sy}. As a consequence, within the perturbative approach these theories represent just an effective description, valid only up to some cutoff scale $\Lambda$, for without a cutoff there would be no scalar self-interactions and without them no symmetry breaking. However, within the variational gaussian approximation it has been  suggested in Ref.~\cite{Consoli:1994jr}  that  this conclusion could not be true. The  point is that the Higgs condensate and its quantum fluctuations could undergo different rescalings when changing the ultraviolet cutoff. Therefore, the relation between $m_H$ and the physical $v_R$ is not the same as in perturbation theory. Indeed, according to this picture   one expects that the condensate rescales as $Z_\varphi \sim \ln \Lambda$ in such a way to compensate the $1/\ln \Lambda$ from $\lambda_R$. As a consequence  the ratio $m_H/v_R$ would  be a cutoff-independent constant. In other words,  one should have:
\begin{equation}
\label{1.1}
m_H \; = \;  \xi \;  v_R  
\end{equation}
where $\xi$ is  a cutoff-independent constant. \\
\indent
It is noteworthy to point out that Eq.~(\ref{1.1}) can be checked by non-perturbative numerical simulations of self-interacting four dimensional scalar field theories on the lattice. Indeed, in previous studies~\cite{Cea:2003gp,Cea:2004ka} we found numerical evidences in support of Eq.~(\ref{1.1}). Moreover, our numerical results showed that the extrapolation to the continuum limit leads to the quite simple result:
\begin{equation}
\label{1.2}
m_H \; \simeq \;  \pi  \;  v_R  \; 
\end{equation}
pointing to  a rather massive Higgs boson without self-interactions (triviality). \\
\indent
The plan of the paper is as follows. In Sec.~\ref{triviality} we illustrate as triviality could coexist with spontaneous symmetry breaking within the simplest self-interacting scalar field theory in four dimensions. In Sec.~\ref{mass} we briefly review the lattice indications  for the non perturbative interpretation of triviality in self-interacting four dimensional scalar field theories and furnish our best numerical determination of the constant $\xi$ in Eq.~(\ref{1.1}). Section~\ref{LHC} is devoted to discuss some experimental signatures of the Higgs boson at LHC.  Finally, our conclusions are drawn in Sec.~\ref{Conclusions}.
\section{Triviality and spontaneous symmetry breaking}
\label{triviality}
In this Section we discuss the triviality and spontaneous symmetry breaking scenario within the simplest scalar field theory, namely a massless real scalar field $\Phi$ with quartic self-interaction $ \lambda \Phi^4$ in four dimensions:
\begin{equation} 
\label{2.1} 
{\cal{L}} =\frac{1}{2}(\partial_\mu \Phi_0)^2
- \frac{1}{4} \lambda_0 \ \, \Phi_0^4 \; ,
\end{equation}
where $\lambda_0$ and $\Phi_0$ are the bare coupling and field respectively. As it is well known~\cite{Coleman:1973,Jackiw:1974}, the one-loop effective potential is given by summing the vacuum diagrams:
\begin{equation} 
\label{2.2} 
V_{1-loop}(\phi_0)  =  \frac{1}{4} \lambda_0 \, \phi_0^4  - \frac{i}{2}  \int \frac{d^4 k}{(2 \pi)^4} \;  \ln  [ - k_0^2 + \vec{k}^2  + 3 \lambda_0  \phi_0^2  
- i \epsilon ] \, .
\end{equation}
Integrating over $k_0$ and discarding a (infinite) constant gives:
\begin{equation} 
\label{2.3} 
V_{1-loop}(\phi_0)  =  \frac{1}{4} \lambda_0 \, \phi_0^4  + \frac{1}{2}  \int \frac{d^3 k}{(2 \pi)^3}  \;  \sqrt{  \vec{k}^2  + 3 \lambda_0  \phi_0^2  } \, .
\end{equation}
This last equation can be interpreted as the vacuum energy of the shifted field:
\begin{equation} 
\label{2.4} 
 \Phi_0 \; = \; \phi_0 \; + \eta \; 
\end{equation}
in the quadratic approximation. Indeed, in this approximation the hamiltonian of the fluctuation $\eta$ over the background $\phi_0$ is:
\begin{equation} 
\label{2.5} 
{\cal{H}}_0  = \frac{1}{2}  (\Pi_{\eta} )^2 \; + \; \frac{1}{2}  (\vec{\nabla} \eta)^2 \; + \; \frac{1}{2}  \; ( 3 \lambda_0  \phi_0^2 ) \; \eta^2 \;  + \; \frac{1}{4} \lambda_0 \, \phi_0^4 \; .
\end{equation}
Introducing an ultraviolet cutoff $\Lambda$ we obtain from Eq.~(\ref{2.3}):
\begin{equation} 
\label{2.6} 
V_{1-loop}(\phi_0)  =  \frac{1}{4} \lambda_0 \, \phi_0^4  + \frac{\omega^4}{64 \pi^2}  \ln \left ( \frac{ \omega^2}{\Lambda^2} \right )   \;  \;  , \; \;  
\omega^2 \, = \,   3 \lambda_0  \phi_0^2 \; .
\end{equation}
It is easy to see that the one-loop effective potential displays a minimum at:
\begin{equation} 
\label{2.7} 
 3 \lambda_0  v_0^2 \; = \;   \frac{\Lambda^2}{\sqrt{e}} \; \exp{ [-   \frac{16 \pi^2}{9 \lambda_0}]} \; .
\end{equation}
Moreover
\begin{equation} 
\label{2.8} 
V_{1-loop}(v_0)   \; = \;  - \;  \frac{\omega^4}{128 \pi^2 } \;  \; ,
\end{equation} 
so that
\begin{equation} 
\label{2.9} 
V_{1-loop}(\phi_0)  \; = \;  \frac{\omega^4}{64 \pi^2} \left [  \ln \left ( \frac{ \phi_0^2}{v_0^2} \right ) - \frac{1}{2} \right ]  \; .
\end{equation}
According to the renormalization group invariance we impose that for $\Lambda \rightarrow \infty$ 
\begin{equation} 
\label{2.10} 
\left [   \Lambda \frac{\partial }{\partial  \Lambda} \; + \; \beta \, \frac{\partial }{\partial  \lambda_0}   \; + \; \gamma \, \phi_0 \,  \frac{\partial }{\partial  \phi_0}
          \right ] V_{1-loop}(\phi_0)  \; = \; 0   \; .
\end{equation}
Within perturbation theory one finds:
\begin{equation} 
\label{2.11} 
\gamma \; = \; 0 \; \; \; \; , \; \; \; \; \beta \;= \; \frac{9}{8 \pi^2} \,  \lambda_0^2 \; .
\end{equation}
Thus, the one-loop corrections have generated spontaneous symmetry breaking. However, the minimum of the effective potential lies outside the expected range of validity of the one-loop approximation and it must be rejected as an artefact of the approximation. On the other hand, as discussed in  Section~\ref{Introduction}, there is no doubt on the triviality of the theory. As a consequence, within perturbation theory there is no room for symmetry breaking. However, following the suggestion of Ref.~\cite{Consoli:1994jr} we argue below that spontaneous symmetry breaking could be compatible with triviality. The arguments go as follows. Write:
\begin{equation} 
\label{2.12} 
 \Phi_0 \; = \; \phi_0 \; + \eta \;  
\end{equation}
where $\phi_0$ is the bare uniform scalar condensate, thus triviality implies that the fluctuation field $\eta$ is a free field with mass $\omega(\phi_0)$. This means that the {\it{ exact}} effective potential is:
\begin{equation} 
\label{2.13} 
V_{eff}(\phi_0)  =  \frac{1}{4} \lambda_0 \, \phi_0^4  + \frac{1}{2}  \int \frac{d^3 k}{(2 \pi)^3}  \;  \sqrt{  \vec{k}^2  + \omega^2(\phi_0)  } \; = \;
 \frac{1}{4} \lambda_0 \, \phi_0^4  + \frac{\omega^4(\phi_0)}{64 \pi^2}  \ln \left ( \frac{ \omega^2(\phi_0)}{\Lambda^2} \right )   \;  \; .
\end{equation}
Moreover, the mechanism of spontaneous symmetry breaking implies that the mass of the fluctuation is related to the scalar condensate as: 
\begin{equation} 
\label{2.14} 
\omega^2(\phi_0) \; = \;  3 \, \tilde{\lambda} \,  \phi_0^2  \;   \; , \; \;  \tilde{\lambda} \; = \; a_1 \, \lambda_0 \; \; ,
\end{equation}
where $a_1$ is some numerical constant.  \\
Now the problem is to see if it exists the continuum limit $\Lambda \rightarrow \infty$.  Obviously, we must have:
\begin{equation} 
\label{2.15} 
\left [   \Lambda \frac{\partial }{\partial  \Lambda} \; + \; \beta(\lambda_0) \, \frac{\partial }{\partial  \lambda_0}   \; + \; \gamma(\lambda_0) \, \phi_0 \,  \frac{\partial }{\partial  \phi_0}
          \right ] V_{eff}(\phi_0)  \; = \; 0   \; .
\end{equation}
Note that now we cannot use perturbation theory to determine $\beta(\lambda_0)$ and $\gamma(\lambda_0)$.  As in the previuos case  the  effective potential displays a minimum at:
\begin{equation} 
\label{2.16} 
 3 \tilde{\lambda}  v_0^2 \; = \;   \frac{\Lambda^2}{\sqrt{e}} \; \exp{ [-   \frac{16 \pi^2}{9 \tilde{\lambda}}]} \; ,
\end{equation}
and
\begin{equation} 
\label{2.17} 
V_{eff}(v_0)   \; = \;  - \;  \frac{m_H^4}{128 \pi^2 } \;  \; , \; \;  m_H^2 \; = \; \omega^2(v_0) \; .
\end{equation} 
Using Eq.~(\ref{2.15}) at the minumum $v_0$ we get:
\begin{equation} 
\label{2.18} 
\left [   \Lambda \frac{\partial }{\partial  \Lambda} \; + \; \beta(\lambda_0) \, \frac{\partial }{\partial  \lambda_0} 
          \right ] m_H^2   \; = \; 0   \; ,
\end{equation} 
which in turns gives:
\begin{equation} 
\label{2.19} 
 \beta(\lambda_0)  \;= \; - \, a_1 \; \frac{9}{8 \pi^2} \,  \tilde{\lambda}^2  \; .
\end{equation}
This last equation implies that the theory is free asymptotically for $\Lambda \rightarrow \infty$ in agreement with triviality:
\begin{equation} 
\label{2.20} 
\tilde{\lambda}  \;  \;  \sim \; \;  \frac{16 \pi^2}{9 a_1} \; \frac{1}{\ln(\frac{\Lambda^2}{m_H^2})}   \; .
\end{equation}
Inserting now Eq.~(\ref{2.19}) into   Eq.~(\ref{2.15}) we obtain:
\begin{equation} 
\label{2.21} 
 \gamma(\lambda_0)  \;= \;  \, a_1^2  \; \frac{9}{16 \pi^2} \,  \tilde{\lambda}  \; .
\end{equation}
This last equation assures that $ \tilde{\lambda} \,  \phi_0^2$ is a renormalization group invariant. Rewriting the effective potential as:
\begin{equation} 
\label{2.22} 
V_{eff}(\phi_0)  =  \frac{(3  \tilde{\lambda} \,  \phi_0^2)^2}{64 \pi^2}    \; \left [
  \ln \left ( \frac{ 3  \tilde{\lambda} \,  \phi_0^2}{m_H^2} \right ) \;  - \; \frac{1}{2}  \right ] \;  \; ,
\end{equation}
we see that $V_{eff}$ is manifestly renormalization group invariant. 

Let us introduce the renormalized field $\eta_R$ and condensate $\phi_R$. Since the fluctuation $\eta$ is a free field we have $\eta_R = \eta$, namely:
\begin{equation} 
\label{2.23} 
Z_{\eta} \;  =  \; 1 \; . 
\end{equation}
On the other hand, for the scalar condensate according to Eq.~(\ref{2.21}) we have:
\begin{equation} 
\label{2.24} 
\phi_R \; = Z_{\phi}^{-\frac{1}{2}} \;  \phi_0 \; \; \; , \; \; Z_{\phi} \; \sim  \;  \lambda_0^{-1} \sim  \;  \ln(\frac{\Lambda}{m_H})  \;  . 
\end{equation}
As a consequence we get that the physical mass $m_H$ is {\it finitely} related to the renormalized vacumm expectation scalar field value $v_R$:
\begin{equation} 
\label{2.25} 
m_H \;  =  \; \xi \; v_R  \; . 
\end{equation}
It should be clear that the physical mass $m_H$ is an arbitrary parameter of the theory (dimensional transmutation). On the other hand the parameter $\xi$ being a pure number can be determined in the non perturbative lattice approach. 

\section{The Higgs boson mass}
\label{mass}
The lattice approach to quantum field theories  offers us the unique opportunity to study a quantum field theory  by means of non perturbative methods. Starting from the classical Lagrangian Eq.(\ref{2.1}) one obtains the lattice theory defined by the Euclidean action:
\begin{equation}
\label{3.1}
S \; = \; \sum_x \left [ \frac{1}{2} \; \sum_{\hat{\mu}} \left ( \Phi(x+\hat{\mu}) -
\Phi(x) \right )^2 \, + \,  \frac{r_0 }{2} \, \Phi^2(x) \, +  \, \frac{\lambda_0}{4} \,
\Phi^4(x) \right ] \; ,
\end{equation}
where $x$ denotes a generic lattice site and, unless otherwise stated,  lattice units are understood.
It is customary to perform numerical simulations in the so-called Ising limit. The Ising limit corresponds
to $\lambda_0 \; \rightarrow \; \infty$. In this limit, the one-component scalar field theory becomes governed by the lattice
action
\begin{equation}
\label{3.2}
S_{\text{Ising}} \;  = \; - \, \kappa  \, \sum_x\sum_{\mu} \,  \left [  \phi(x+\hat
e_{\mu}) \phi(x) \, + \, \phi(x-\hat e_{\mu}) \phi(x)  \right ]
\end{equation}
with $\Phi(x)=\sqrt{2\kappa}\, \phi(x)$ and where $\phi(x)$ takes only the 
values $+1$ or $-1$.  
\\
It is known that there is a critical coupling~\cite{Gaunt:1979aa}:
\begin{equation}
\label{3.3}
\kappa_c \; = \;   0.074834(15) \; ,
 \end{equation}
such that for $\kappa > \kappa_c$ the theory is in the  broken phase, while for  $\kappa <  \kappa_c$ it
is in the symmetric phase. The continuum limit corresponds to $\kappa \rightarrow \kappa_c$ where $m_{latt} \equiv a m_H \rightarrow 0$, $a$ being the lattice spacing. \\
As discussed in  Section~\ref{Introduction}, the triviality of the scalar theory means that the renormalized self coupling vanishes as   $\frac{1}{\ln(\frac{\Lambda^2}{m_H^2})}$ when   $\Lambda \rightarrow \infty$. As a consequence in the continuum limit the theory admits a gaussian fixed point. 
\\
On the lattice the ultraviolet cutoff is $\Lambda = \frac{\pi}{a}$ so that we have:
\begin{equation} 
\label{3.4-a} 
 \lambda \; \sim  \; \frac{1}{ \ln(\frac{\Lambda}{m_H})}  \; \sim  \; \frac{1}{ \ln(\frac{\pi}{a m_H})} \; = \;  \frac{1}{ \ln(\frac{\pi}{m_{latt}})} \; . 
\end{equation}
The perturbative interpretation of triviality~\cite{Luscher:1987ay,Luscher:1988ek} assumes that in the continuum limit there is an {\it{infrared}} gaussian fixed point where the limit $m_{latt} \rightarrow 0$ corresponds to  $m_{H} \rightarrow 0$. On the other hand,  according to Section~\ref{triviality},  in the triviality and spontaneous symmetry breaking scenario the continuum dynamics is governed by an  {\it{ultraviolet}} gaussian fixed point  where $m_{latt} \rightarrow 0$ corresponds to  $a \rightarrow 0$.  As we discuss below, these two different interpretation of triviality lead to different logarithmic correction to the gaussian scaling laws which can be checked with numerical simulations on the lattice. \\
\indent
In  Ref.~\cite{Cea:2004ka} extensive numerical lattice simulations of the  one-component scalar field theory in the Ising limit 
have been  performed. In particular,
using the Swendsen-Wang~\cite{Swendsen:1987ce}  and Wolff~\cite{Wolff:1989uh} cluster algorithms 
the bare magnetization (vacuum expection value):
\begin{equation}
\label{3.4}
 v_{latt} \; = \; \langle |\phi| \rangle \quad , \quad \phi \equiv \frac{1}{L^4} \; \sum_x
\phi(x)
\end{equation}
and the bare zero-momentum susceptibility:
\begin{equation}
\label{3.5}
 \chi_{{latt}}=L^4 \left[ \left\langle |\phi|^2
\right\rangle - \left\langle |\phi| \right\rangle^2 \right] 
\end{equation}
have been computed.
%
%
\begin{figure}[t]
\includegraphics[width=0.9\textwidth,clip]{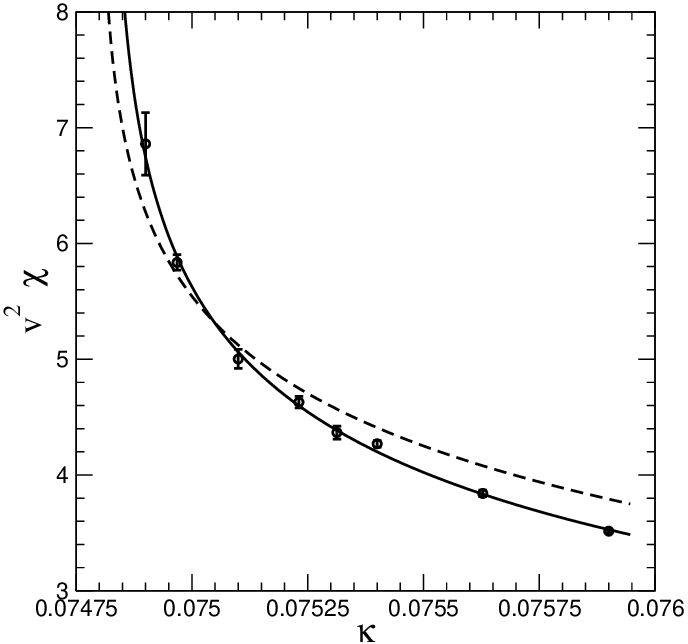}
\caption{\label{fig.1} We show the lattice data for $ v_{latt}^2 \;  \chi_{latt}$
together with the fit Eq.~(\ref{3.7}) (solid line) and the two-loop fit Eq.~(\ref{3.9}) (dashed line)
where the fit parameters $a_1$ and $a_2$ are allowed to vary inside their  theoretical uncertainties
Eq.~ (\ref{3.10}).}
\end{figure}
According  to the perturbative scheme of Refs.~\cite{Luscher:1987ay,Luscher:1988ek} one expects
\begin{equation}
\label{3.5-a}
      v_{latt}^2 \;  \chi_{latt}  \; \sim
|\ln(\kappa-\kappa_c)| \; , \; \kappa \rightarrow \kappa_c^{+}
\end{equation}
On the other hand, since in the triviality and spontaneous symmetry breaking scenario one expects that
$Z_\varphi  \sim \ln(\frac{\Lambda}{m_H})  \sim|\ln(\kappa-\kappa_c)| $ we have:
\begin{equation}
\label{3.6}
       v_{latt}^2 \;  \chi_{latt}  \sim |\ln(\kappa-\kappa_c)|^2 \; , \; \kappa \rightarrow \kappa_c^{+} \; .
       \end{equation}
The predictions in Eq.~(\ref{3.6})  can be directly compared with the lattice data  reported
in  Ref.~\cite{Cea:2004ka} and displayed in Fig.~\ref{fig.1}. We fitted the data to the 2-parameter form:
\begin{equation}
\label{3.7}
 v_{latt}^2 \;  \chi_{latt} \; = \; \alpha |\ln(\kappa-\kappa_c)|^2 \; .
\end{equation}
We obtain a rather good fit of the lattice data (full line in Fig.~\ref{fig.1}) with
\begin{equation}
\label{3.8}
  \alpha \; = \;  0.07560(49)  \; \; , \; \;   \kappa_c \; = \; 0.074821(12) \; \; , \; \; \chi^2_{dof} \simeq  1.5 \; .
  \end{equation}
Note that  our precise determinations of the critical coupling $\kappa_c$ in Eq.~(\ref{3.8}) is in good
agreement with the value obtained in Ref.~\cite{Gaunt:1979aa} (see Eq.~(\ref{3.3})~).\\
On the other hand,  the prediction based on 2-loop renormalized perturbation theory is~\cite{Luscher:1988ek,Balog:2004zd}
($l=|\ln (\kappa -\kappa_c)|$):
\begin{equation}
\label{3.9}
      \left[  v_{latt}^2 \;  \chi_{latt}  \right]_{\text{2-loop}}=
a_1( {l} - \frac{25}{27} \ln { l}) +a_2
\end{equation}
together with the theoretical relations:
\begin{equation}
\label{3.10}
a_1\; = \; 1.20(3) \; \; \; \; , \; \; \;  a_2 \; = \; - \, 1.6(5) \; \; .
\end{equation}
We fitted the lattice data to Eq.~(\ref{3.9})   by allowing the fit parameters $a_1$ and $a_2$
to vary inside their  theoretical uncertainties Eq.~(\ref{3.10}). The fit resulted in  (dashed line in Fig.~\ref{fig.1}):
\begin{equation}
\label{3.11}
  a_1 \; = \;  1.17  \; , \;   a_2 \; = \; - \,  2.10  \; \; , \; \; \kappa_c \; = \; 0.074800(1) \; \; , \; \; \chi^2_{dof} \simeq  132 \; .
\end{equation}
It is evident from   Fig.~\ref{fig.1} that the quality of the 2-loop fit is poor. 
However, these results have been criticized  by the authors of
Ref.~\cite{Balog:2004zd} and have given rise to an intense debate in the recent literature~\cite{Cea:2005ad,Stevenson:2005yn,Balog:2006fs,Castorina:2007ng,Lundow:2009}.
\\
\indent
Additional numerical evidences would come from the direct detection of the condensate rescaling $Z_\phi   \sim |\ln(\kappa-\kappa_c)| $
on the lattice. To this end, we note that:
\begin{equation}
\label{3.12}
  Z_\phi  \; \equiv   \;   2 \; \kappa m^2_{latt}  \; \chi_{latt} \; .
\end{equation}
In Fig.~\ref{fig.2} we display the lattice data  obtained in Ref.~\cite{Cea:2004ka}  for $Z_\phi$, as defined in
Eq.~(\ref{3.12}) versus $m_{latt}$  reported in  Ref.~\cite{Luscher:1988ek} at the various
values of $\kappa$.  For comparison we also report the perturbative prediction of $Z_{\eta}$ taken from   Ref.~\cite{Luscher:1988ek}.
We try to fit the lattice data with:
\begin{equation}
\label{3.13}
 Z_\phi \;   = \; A \;  \ln ( \, \frac{\pi}{m_{latt}} \, ) \; .
\end{equation}
Indeed, we obtain a satisfying fit to the lattice data (solid line in Fig.~\ref{fig.2}):
\begin{equation}
\label{3.14}
  A \; = \;  0.498(5)  \; , \;   \; \;  \chi^2_{dof} \simeq  4.1  \; .
\end{equation}
\indent
%
%
%
%
\begin{figure}[t]
\includegraphics[width=0.9\textwidth,clip]{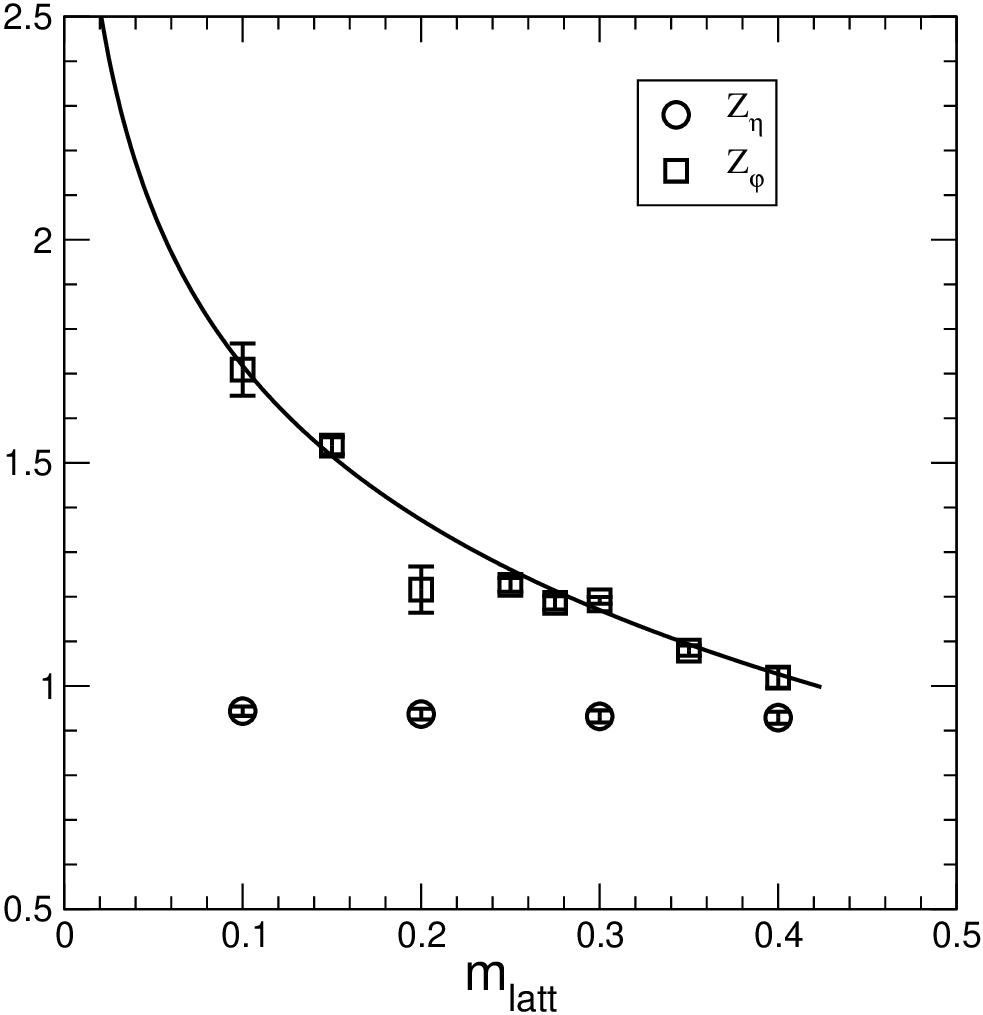}
\caption{\label{fig.2} The lattice data for $Z_\phi$, as defined in
Eq.~(\ref{3.12}), and the perturbative prediction $Z_{\eta}$
versus  $m_{latt} $. The solid line is the fit to  Eq.~(\ref{3.13}).
}
\end{figure}
By adopting this alternative interpretation of triviality there are important phenomenological implications. In fact,  assuming to know the value of $v_R$,  the ratio $\xi \; = \; m_H/v_R$ is now a cutoff-independent quantity. Indeed,  the physical $v_R$ has to be computed from the bare $v_B$
 through $Z=Z_\varphi$  rather than through the perturbative $Z=Z_{\eta}$. In this case the perturbative relation~\cite{Luscher:1988ek}:
\begin{equation}
\label{3.14-a}
  \frac{m_H}{v_R} \;  \equiv \; \sqrt {  \frac{\lambda_R}{3} } 
\end{equation}
becomes
\begin{equation}
\label{3.15}
 \frac{m_H}{v_R} \; = \;  \sqrt{ \frac{\lambda_R}{3} \, \frac{Z_\varphi}{Z_{\eta} } }
 \;  \equiv \; \xi
\end{equation}
obtained by replacing $Z_{\eta}$ with $Z_{\varphi}$ in Ref.~\cite{Luscher:1988ek} and correcting for the perturbative $Z_{\eta}$.
Using the values of $\lambda_R$ reported in  Ref.~\cite{Luscher:1988ek} and our values of $Z_\varphi$, we display in Fig.~\ref{fig.3}
the values of $m_H$ as defined through  Eq.~(\ref{3.15}) versus $m_{latt}$ for $v_R \; =\; 246$ {\text{GeV}}. The error band corresponds to a one standard deviation error in the determination of $m_H$ through a fit with a constant function. As one can see, the $Z_\varphi \sim \ln \Lambda$ trend observed
in Fig.~\ref{fig.2} compensates the $1/\ln \Lambda$ from $\lambda_R$ so that $\xi$ turns out to be a cutoff-independent constant:
\begin{equation}
\label{3.16}
\xi \; = \; 3.065(80) \; \; \; , \; \; \; \chi^2_{dof} \simeq  3.0 \; ,
\end{equation}
which corresponds to:
\begin{equation}
\label{3.17}
m_H \; =754 \; \pm 20 \; \pm 20 \;  \; {\text{GeV}}
\end{equation}
where the last error is our estimate of systematic effects. \\
\begin{figure}[t]
\includegraphics[width=0.9\textwidth,clip]{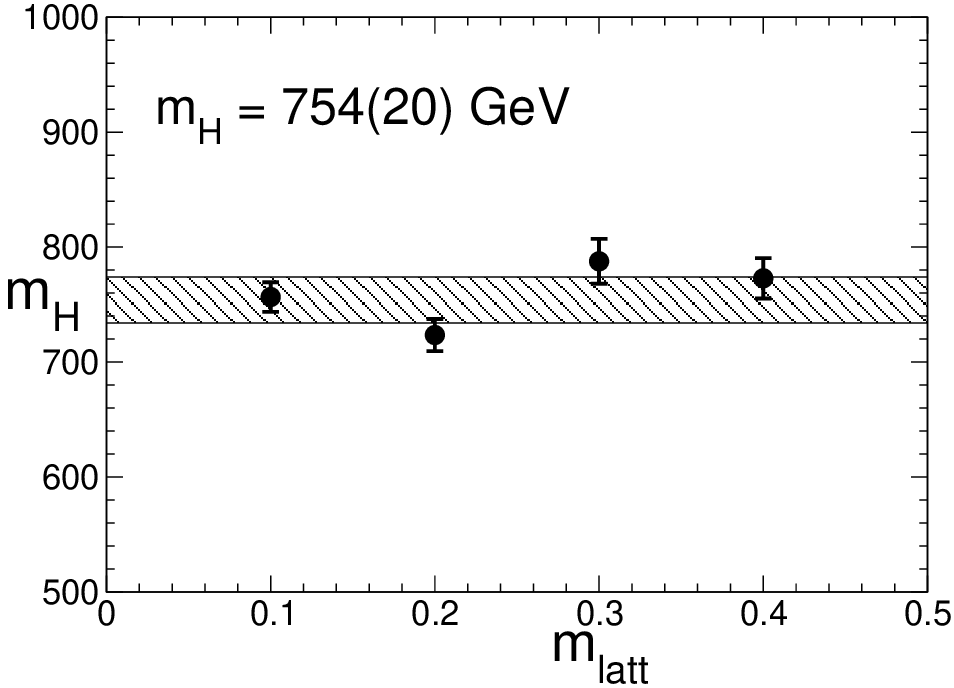}
\caption{\label{fig.3} The values of $m_H$ as defined through Eq.~(\ref{3.15}) versus
$m_{latt}$ assuming $v_R \; =\; 246$ GeV. The error band corresponds to  one standard deviation
error in the determination of $m_H$ through a fit with a constant function.}
\end{figure}
One could object that our lattice estimate of the Higgs mass Eq.~(\ref{3.17}) is not relevant for the physical Higgs boson. Indeed, the scalar theory relevant for the Standard Model is the O(4)-symmetric self-interacting theory. However, the Higgs mechanism eliminates three scalar fields leaving as physical Higgs field  the radial excitation whose dynamics  is described by  the one-component self-interacting scalar field theory. Therefore, we are confident that our determination of  the Higgs mass applies also to the Standard Model Higgs boson.
\section{The Higgs physics at LHC}
\label{LHC}
Recently, both the ATLAS and CMS collaborations~\cite{Nisati:2011,Sharma:2011} reported the experimental results for the search of the  Higgs boson at the Large Hadron Collider
running at $\sqrt{s} = 7$ TeV, based on a total integrated luminosity between 1 fb$^{-1}$ and  2.3 fb$^{-1}$. \\
It is worthwhile to briefly discuss the main physical properties of our proposal for the trivial Higgs boson.
For Higgs mass in the range $700 - 800 \; {\text{GeV}}$ the main production mechanism at LHC is the gluon fusion $gg  \rightarrow H$. The theoretical estimate of the production cross section at LHC for centre of mass  energy $\sqrt{s} = 7 \, {\text{TeV}}$ is~\cite{Dittmaier:2011ti} :
\begin{equation}
\label{4.1}
\sigma(gg  \rightarrow H) \; \simeq \; 0.06  -  0.14  \; {\text{pb}} \; \; , \; \;  700 \; {\text{GeV}} \; < m_H \; < \; 800   \;  {\text{GeV}} \; .
\end{equation}
The gluon coupling to the Higgs boson in the Standard Model is mediated by triangular loops of top and bottom quarks. Since the Yukawa coupling of the Higgs particle to heavy quarks grows with quark mass, thus bilancing the decrease of the triangle amplitude, the effective gluon coupling approaches a non-zero value for large loop-quark masses. On the other hand, we argued that the Higgs condensate rescales with $Z_{\phi}$. This means that, if the fermions acquires a finite mass through the Yukawa couplings, then we are led to conclude that the coupling of the physical Higgs field to the fermions could be very different from the Standard Model Higgs boson.  On the other hand, the coupling of the Higgs field to
the gauge vector bosons is fixed by the gauge symmetries. So that the coupling of our Higgs boson to the  gauge vector bosons is the same as for the Standard Model Higgs boson.  
For large Higgs masses the vector-boson fusion mechanism becomes competitive to gluon fusion Higgs production~\cite{Dittmaier:2011ti}:
\begin{equation}
\label{4.2}
\sigma(W^+ \, W^-   \rightarrow H) \; \simeq \; 0.02  -  0.03  \; {\text{pb}} \; \; , \; \;  700 \; {\text{GeV}} \; < m_H \; < \; 800   \;  {\text{GeV}} \; .
\end{equation}
\indent
The main difficulty in the experimental identification of a very heavy Standard Model Higgs ($m_H > 650 \; {\text{GeV}}$)  resides in the  large width which makes impossible to observe a mass peak.  However, in the triviality and spontaneous symmetry breaking scenario the Higgs self-coupling vanishes so that the decay width is mainly given by the  decays  into pairs of massive gauge bosons. Since the Higgs is trivial there are no loop corrections due to the Higgs self-coupling and we obtain for the Higgs total width:
\begin{equation}
\label{4.3}
\Gamma( H)  \; \simeq \; \Gamma( H \rightarrow W^+ \, W^-)  \; + \; \Gamma( H \rightarrow Z^0 \, Z^0) \; 
\end{equation}
where~\cite{Higgs_book,Djouadi:2005gi}
\begin{equation}
\label{4.4}
 \Gamma( H \rightarrow W^+ \, W^-)   \; \simeq \;  \frac{G_F m_H^3}{8 \sqrt{2 \pi}} \, \sqrt{1 - 4 x_W} \, (1 \, - \, 4 x_W \, + 12 x_W^2) \; , \; x_W \; = \; \frac{m_W^2}{m_H^2} 
\end{equation}
\begin{equation}
\label{4.5}
 \Gamma( H \rightarrow Z^0 \, Z^0)   \; \simeq \;  \frac{G_F m_H^3}{16 \sqrt{2 \pi}} \, \sqrt{1 - 4 x_Z} \, (1 \, - \, 4 x_Z \, + 12 x_Z^2) \; , \; x_Z \; = \; \frac{m_Z^2}{m_H^2} \; . 
\end{equation}
Assuming $m_H \, \simeq \, 750 \; {\text{GeV}}$,  $m_W \, \simeq \, 80 \; {\text{GeV}}$ and $m_Z \, \simeq \, 91 \; {\text{GeV}}$, we obtain:
\begin{equation}
\label{4.6}
\Gamma( H)  \; \simeq \; 340 \; {\text{GeV}} \; . 
\end{equation}
\indent
A thorough discussion of  the experimental signatures of our Trivial Higgs is presented in Ref.~\cite{Cea:2011} where
we compare our proposal with the recent data from ATLAS and CMS collaborations  based on a total integrated luminosity between 1 fb$^{-1}$ and 
 2.3 fb$^{-1}$. In fact, we argue that the available experimental data
seem to  be consistent with our scenario.
\section{Conclusions}
\label{Conclusions}
The Standard Model  requires the existence of a scalar Higgs boson to break electroweak symmetry and provide mass terms to gauge bosons and fermion fields. Usually the spontaneous symmetry breaking in the Standard Model is implemented within the perturbation theory which leads  to predict that the Higgs boson mass squared is proportional to the self-coupling. However, there exist several results which point to vanishing scalar self-coupling.  Therefore, within the perturbative approach scalar field  theories represent just an effective description valid only up to some cutoff scale, for without a cutoff there would be no scalar self-interactions and without them no symmetry breaking.  In other words, spontaneous symmetry breaking is incompatible with strictly local scalar fields in the perturbative approach. \\
\indent
In this paper we have shown that  local scalar fields are compatible with spontaneous symmetry breaking. In this case, the continuum dynamics is governed by an ultraviolet gaussian fixed point (triviality) and a non-trivial rescaling of the scalar condensate. We argued that non-perturbative lattice simulations are consistent with this scenario. Moreover, we find that  the  Higgs boson is rather heavy. Finally, the non-trivial rescaling of the Higgs condensate suggests that the whole issue of  generation of fermion masses through the Yukawa couplings must be reconsidered.
%
%
%

\end{document}